\documentclass{aastex631}

\usepackage{amsmath}
\usepackage{graphicx}	% Including figure 
\usepackage{rotating}

\begin{document}

\title[Kicks and Spins]{A Theory for Neutron Star and Black Hole Kicks and Induced Spins}

\author[0000-0002-3099-5024]{Adam Burrows}
\affiliation{Department of Astrophysical Sciences, Princeton University, Princeton NJ 08544 and the Institute for Advanced Study, 1 Einstein Drive, Princeton NJ  08540}

\correspondingauthor{Adam Burrows}
\email{aburrows@princeton.edu}

\author[0000-0002-0042-9873]{Tianshu Wang}
\affiliation{Department of Astrophysical Sciences, Princeton University, Princeton, NJ 08544}

\author[0000-0003-1938-9282]{David Vartanyan}
\affiliation{Carnegie Observatories, 813 Santa Barbara St., Pasadena, CA 91101}

\author[0000-0001-5939-5957]{Matthew S.B. Coleman}
\affiliation{Department of Physics and Engineering Physics, Stevens Institute of Technology, Hoboken, NJ, 07030}

%matthew.s.b.coleman@gmail.com

\date{Accepted XXX. Received YYY}

\begin{abstract}
Using twenty long-term 3D core-collapse supernova simulations, we find that lower compactness progenitors that explode quasi-spherically due to the short delay to explosion experience smaller neutron star recoil kicks in the $\sim$100$-$200 km s$^{-1}$ range, while higher compactness progenitors that explode later and more aspherically leave neutron stars with kicks in the $\sim$300$-$1000 km s$^{-1}$ range. In addition, we find that these two classes are correlated with the gravitational mass of the neutron star. This correlation suggests that the survival of binary neutron star systems may in part be due to their lower kick speeds. We also find a correlation of the kick with both the mass dipole of the ejecta and the explosion energy. Furthermore, one channel of black hole birth leaves masses of $\sim$10 $M_{\odot}$, is not accompanied by a neutrino-driven explosion, and experiences small kicks. A second is through a vigorous explosion that leaves behind a black hole with a mass of $\sim$3.0 $M_{\odot}$ kicked to high speeds.  We find that the induced spins of nascent neutron stars range from seconds to $\sim$10 milliseconds, {but do not yet see a significant spin/kick correlation for pulsars.} We suggest that if an initial spin biases the explosion direction, a spin/kick correlation {would be} a common byproduct of the neutrino mechanism of core-collapse supernovae.  Finally, the induced spin in explosive black hole formation is likely large and in the collapsar range. This new 3D model suite provides a greatly expanded perspective and appears to explain some observed pulsar properties by default. 
\end{abstract}

%\keywords{Supernova, Nucleosynthesis, R-process}

%\begin{keywords}
\keywords{(stars:) supernovae: general -- (stars:) neutron -- (stars:) pulsar -- neutrinos  -- hydrodynamics}
%\end{keywords}

%%%%%%%%%%%%%%%%%%%%%%%%%%%%%%%%%%%%%%%%%%%%%%%%%%

%%%%%%%%%%%%%%%%% BODY OF PAPER %%%%%%%%%%%%%%%%%%

\section{Introduction}
\label{intro}

Radio pulsars \citep{Manchester2005,Kaspi2016}, magnetars \citep{1998Natur.393..235K,Kaspi2017}, and low-mass and high-mass X-ray binaries (LMXBs and HMXBs) \citep{Shakura,Remillard} are or contain neutron stars or black holes \citep{macleod} created at the death of a massive star, oftimes in a core-collapse supernova explosion. Excitingly, the mergers of tight binaries of stellar-mass black holes and neutron stars have recently been captured as gravitational-wave sources \citep{Ligo}. Interestingly, a broad spectrum of observations indicates that the formation of neutron stars oftimes leaves them with significant kick speeds that range up to  $\sim$1500 km s$^{-1}$, with an average kick speed of $\sim$350-400 km s$^{-1}$ \citep{Faucher2006,Chatterjee2009}
and possible spin-kick correlations \citep{Holland-Ashford2017,2018ApJ...856...18K,Ng2007}. However, the bound neutron star population in globular clusters suggests that some neutron stars are born with kick speeds below $\sim$50 km s$^{-1}$ \citep{Lyne1994,Arzoumanian2002}, consistent with the suggestion that the kick distribution could be bimodal \citep{Cordes1998,Arzoumanian2002,Verbunt2017}. In contradistinction, studies of black holes in X-ray binaries imply average kick speeds for them of less than $\sim$50 km s$^{-1}$ \citep{Fragos2009,2023_gomez_bh,macleod}, but with a few possible exceptions \citep{Mandel2016,Fragos2009,Repetto2017,atri,dashwood_brown}. 

These data and inferences require explanation in the context of compact object birth, i.e. in the context of core-collapse supernova explosions (CCSNe). The Blaauw mechanism due to unbinding of progenitor binaries as a result of explosive mass loss \citep{Blaauw,Hills,Blaauw2}
is quantitatively inadequate, though often subdominantly operative. Models that invoke asymmetrical radio pulsar winds \citep{Harrison1975} are not compelling \citep{Lai2001}. Models that rely on neutrino jets in the context of rapid rotation at birth \citep{Spruit1998} are not consistent with modern CCSN simulations \citep{Janka2017,Stockinger2020,Burrows2020,Burrows2021,coleman}. The most comprehensive and well-developed theory for pulsar and black hole kicks at birth involves recoil due to the generically asymmetrical supernova explosion \citep{Burrows1996,Scheck2006,Burrows2007,Nordhaus2010,Nordhaus2012,Wongwathanarat2013,Janka2017,Gessner2018,muller_low_kick2019,Nakamura2019,coleman} and to the accompanying aspherical neutrino emissions \citep{2006ApJS..163..335F, Nagakura2019b, rahman2022}, the latter generally being subdominant for neutron star birth but important for black hole birth \citep{coleman}. 

{Asphericities arise in various regions of the supernova progenitor star and the timescales of their effects can vary from a few seconds to hours or days. Simultaneous accretion and explosion leads to interactions between ejecta and accreta, and this phenomenon is the first major physical process that influences the kick velocity experienced by the central object. When accretion terminates and the explosive shock wave has swept away most of the outer envelope a few seconds later, this process ceases and the core has achieved its final asymptotic kick state. This marks the end of kick evolution if the explosion is sufficiently energetic, i.e., if the total energy of the outgoing matter is higher than the binding energy of the outer envelope. For less energetic explosions, part of the ejected matter may slow down due to interactions with stellar envelopes, and may fall back onto the central object \citep{schroder2018,Chan2020}. This second process may last hours to days. Even if the progenitor fails to explode and there is no asymmetric ejecta, the asymmetric neutrino emission can still provide a modest kick \citep{2006ApJS..163..335F, Nagakura2019b, rahman2022}. In this case, the accretion of the outer envelope can carry angular momenta to the compact core \citep{antoni_2023}.}

{In this paper, we focus on the first process mentioned above, i.e., kicks during the simultaneous accretion and explosion phase. This is the dominant process in energetic CCSN explosions \citep{coleman,Chan2020}. Many relevant simulations have been done in the past \citep{Burrows1996,Scheck2006,Burrows2007,Nordhaus2010,Nordhaus2012,Wongwathanarat2013,Janka2017,Gessner2018,muller_low_kick2019,Nakamura2019,coleman}, but} the required sophisticated and expensive 3D simulations have not been continued to late enough times after bounce to achieve asymptotic kick speeds, with very few exceptions \citep{Stockinger2020,coleman}. Most 3D simulations are stopped within one second or less after bounce, far to soon to witness the final kick speed imparted. Moreover, even the exceptional longer-term 3D studies did not explore the systematics with the broad range of progenitor masses to determine the correlations between kick speed and progenitor mass and/or initial core structure, the latter frequently associated with the compactness parameter \footnote{The compactness parameter is defined to be $\frac{M/M_\odot}{R(M)/1000\text{km}}$, for which we generally prefer to set $M = 1.75 M_{\odot}$. The \citet{oconnor2011} paper which originated the use of compactness did so in the context of black hole formation and used a value near the maximum neutron star baryon mass of 2.5 $M_{\odot}$, which is less appropriate in the supernova context.}. They focused on the few models whose recoil kick had quickly settled to nearly a final state. {In contrast to papers incorporating detailed numerical simulation, papers such as \citet{Janka2017} and \citet{Mandel2020} and \citet{bray2016} and \citet{bray2018} have derived semi-analytical and/or empirical relations between kick velocities and supernova/progenitor properties. Although such simple recipes are intuitively appealing, they need to be tested thoroughly using 3D numerical simulations.}

Hence, a comprehensive theoretical study of the broad spectrum of systematic behaviors in the context of modern 3D core-collapse supernova theory has not yet been performed.  With this paper, we attempt to remedy this situation by providing and analyzing twenty of our recent long-term state-of-the-art 3D F{\sc{ornax}} \citep{Skinner2019} simulations for initially non-rotating ZAMS (zero-age main-sequence star) solar-metallicity progenitors from 9 to 60 $M_{\odot}$ taken from \citet{Sukhbold2016} and \citet{Sukhbold2018}. {Our goal is to build a more solid foundation for the current CCSN kick theory and to provide detailed 3D models that can be used to test, calibrate, and update widely-used recipes.}  

{We emphasize that we are addressing in this paper only solar-metallicity progenitors and those most likely to be the origin of most galactic neutron stars and black holes and leave sub-solar metallicity studies to future work. We also are not here addressing the issue of the potential role of very massive massive stars and pulsational pair instabilities\footnote{The reader might find \citet{rahman2022} useful in this regard.}.} Importantly, many of our models have been carried in 3D beyond four seconds after bounce and show signs of asymptoting to their final recoil kick speeds. A few of these models have or will form black holes. 

In this paper, we complement our kick study with a study of the associated induced spins \citep{Blondin2007a,Rantsiou2011,coleman}. Though a compelling explanation for pulsar and neutron star spins is spin up from an initially spinning Chandrasekhar core upon collapse, modern supernova simulations leave proto-neutron star (PNS) cores spinning due to random and stochastic post-explosion accretion of angular momentum ($L$), even when the core is initially non-rotating. In fact, the magnitude of such induced spins \citep{coleman} can be comparable to measured pulsar spins \citep{Chevalier1986,Lyne1994,Manchester2005,Faucher2006,Popov2012,Igoshev2013,Noutsos2013}. Moreover, there are hints that radio pulsar spins and kicks might be correlated at birth \citep{Wang2006,Wang2007,Johnston2005,Johnston2007,Ng2007,Noutsos2013}
Finally, angular momentum transport from the stellar core to the mantle up to the time of core collapse might perhaps be more efficient than hitherto thought \citep{Cantiello}, leaving the Chandrasekhar core slowly rotating just before implosion.  Therefore, we continue here the exploration of the possibility that pulsar spins might be partially reflective of induced spins during the supernova. Similarly, black holes created in supernova explosions \citep{Burrows2023} will have both large kicks and significant induced spin parameters ($a = Lc/GM^2$), though black holes created in a non-exploding context should have birth spins reflective of the initial progenitor spin (though perhaps in a complicated fashion). {However, similar to the case with kicks, the induced spins of the central object can be influenced by different very long-term processes. In this paper, however, we focus on the crucial first few seconds post-bounce.} 

We used the code F{\sc{ornax}} \citep{Skinner2019,Radice2019,Nagakura2019, Nagakura2019b,Vartanyan2019b,Burrows2019,Burrows2020,Vartanyan2020,Nagakura2020,Vartanyan2021,Vartanyan2023,Burrows2023} to generate the twenty 3D models to late times after bounce used here to study recoil kicks and induced spins in the context both of supernova explosions and quiescent black hole formation.
A more comprehensive set of studies using these new 3D simulations is in preparation which will address other aspects and outcomes of core-collapse and their dependence upon progenitor. This is the largest set of long-term (many seconds after bounce) 3D state-of-the-art core-collapse simulations ever created and we hope to use it to gain new insights into the core-collapse supernova phenomenon. To create this set, we employed the CPU machines TACC/Frontera \citep{Stanzione2020} and ALCF/Theta and the GPU machines ALCF/Polaris and NERSC/Perlmutter. The specific setups and run parameters are described in \citet{Burrows2023} and \citet{Vartanyan2023}. The techniques employed to derive the kicks and induced spins are described in detail in \citet{coleman} and follow the standard approaches in the literature \citep{Scheck2006,Wongwathanarat2013,Stockinger2020}. However, for the calculation of the neutrino component, unlike in the \citet{coleman} paper, we calculate its contribution here at the PNS radius (defined as the 10$^{11}$ g cm$^{-3}$ isosurface), and not at the outer mass surface of the final proto-neutron star (PNS). The result for the final total vector kick is the same, but there is a different partitioning between the neutrino and matter components. The momentum transfer between the matter and the radiation in the region between these two definitions accounts for the shift and the total momentum is conserved under both approaches.  We employ the different method here to comport with the standard practice in the literature \citep{Stockinger2020}.

\section{Kicks to Neutron Stars at Birth}
\label{kicks}

Until recently, 3D models experiencing kicks had been calculated to post-bounce times insufficient to actually witness their final values.  Moreover, due to computational limitations, the systematics with progenitor core structure and total ZAMS mass were obscured, since only one or a few models early during the accumulating recoil could be simulated.  This is akin to viewing the entire progenitor panorama through a straw.  The stochasticity of the final values of any explosion observable inherent in the chaotic turbulence of the CCSN phenomenon also served to compromise extrapolation and interpretation.  However, with our new set of twenty long-term 3D full-physics simulations, we are now able to extract what we suggest are real trends across the progenitor continuum and find correlations among observables 
hitherto only imperfectly glimpsed. Hence, though for all core-collapse supernova observables there are expected to be a distribution of values, even for the same progenitor, the availability of a raft of such models across the progenitor continuum mitigates in part against this modest impediment to determining system-wide trends.  
This is the philosophy of our multi-model study, enabled by the efficiency of F{\sc{ornax}} and the availability of 
high-performance computing resources.

\begin{figure}
    \centering
    \includegraphics[width=0.48\textwidth]{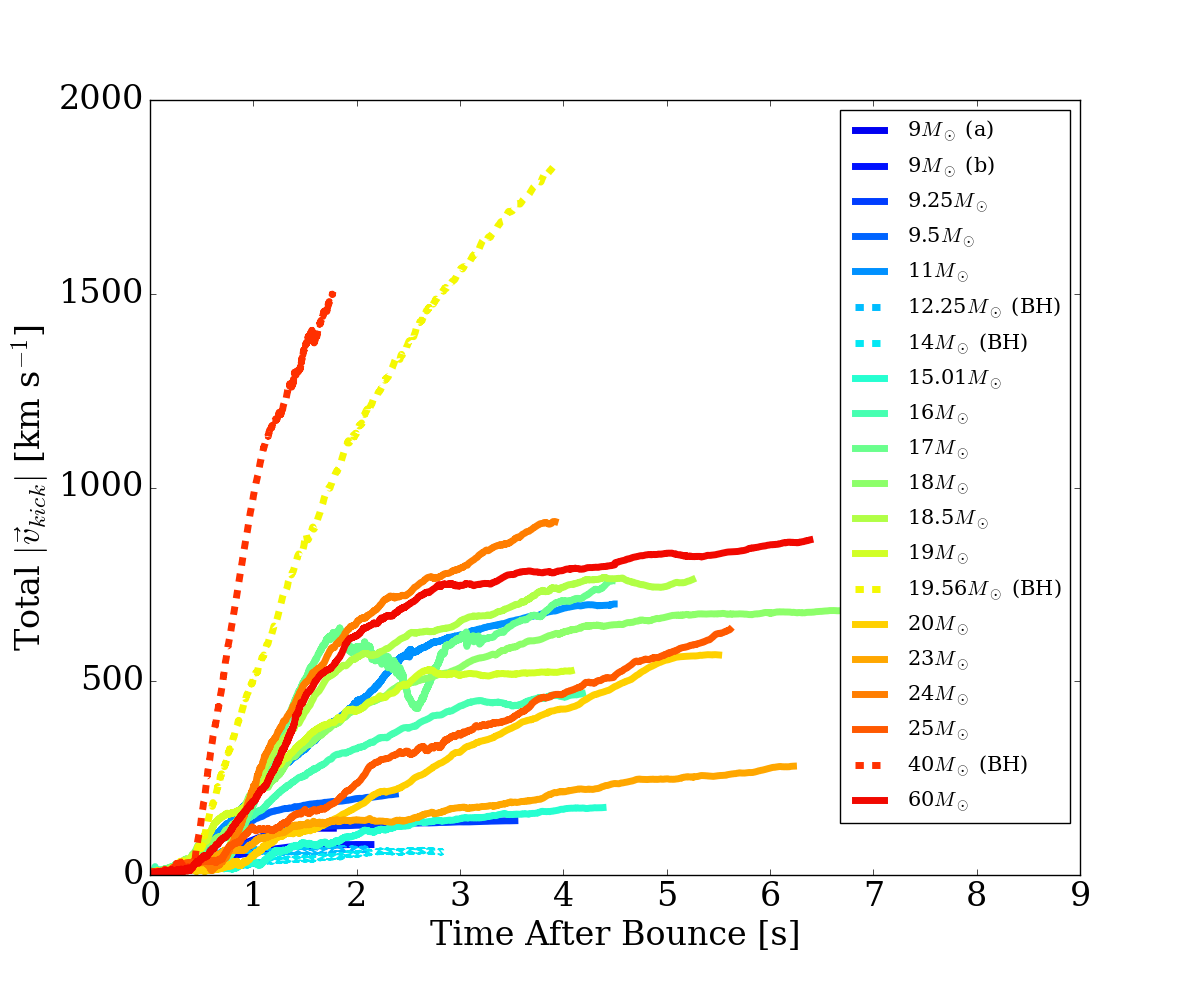}
    \includegraphics[width=0.48\textwidth]{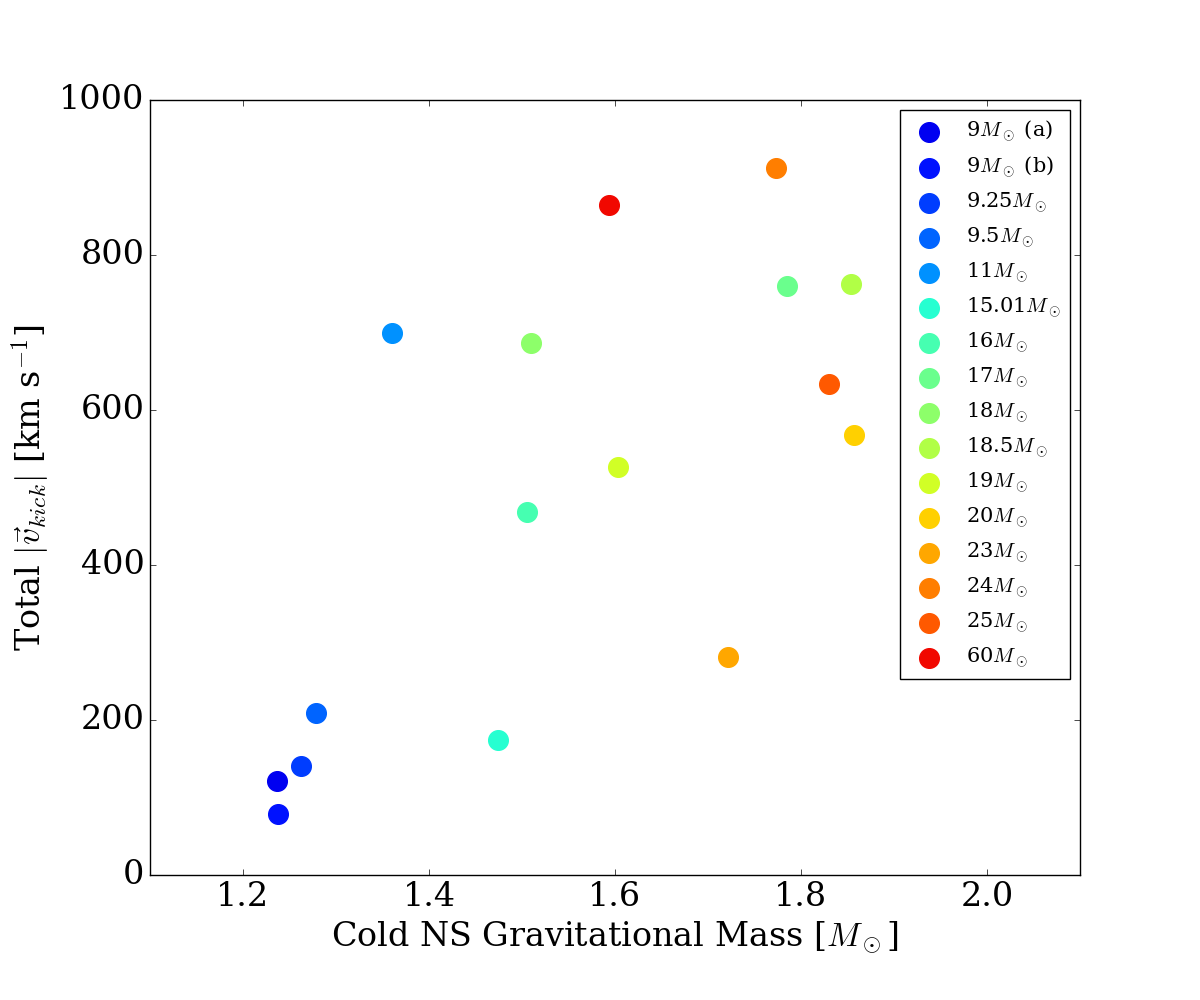}
    \caption{{\bf Left:} Total kick velocity as a function of time in each simulation. Solid lines are models that form neutron stars, while dashed lines are black hole formers. {\bf Right:} Kick versus cold NS gravitational mass. Black hole formers are not included here.}
    \label{fig:total-kick-t}
\end{figure}

Table \ref{tab:sims} provides summary numbers of relevance to this study.  The top rank displays the data for the exploding models that leave neutron stars (NS), while the bottom rank displays the corresponding data for the models that leave black holes (BH) (see \S\ref{black}), either quiescently or explosively. Most of the models have either 
asymptoted to the final state (in particular the lowest-mass,
lowest-compactness progenitors) or nearly so. The left-hand side of Figure \ref{fig:total-kick-t} depicts the temporal evolution of the absolute value of the vector kick, including both the matter and neutrino recoils and all momentum flux, pressure, and gravitational forces \citep{coleman}. The lowest-mass
progenitors generally flatten/asymptote earliest, while the models with the largest compactness (frequently more massive) take longer. Such long-duration 3D simulations have not in the past been available.

The right-hand side of Figure \ref{fig:total-kick-t} depicts the final kicks versus final gravitational mass of the corresponding neutron stars.  This is a plot of observables that has not been available before and speaks to the importance of long-term 3D simulations to provide testable predictions for modern CCSN theory. We note that the lowest mass progenitors (blue) are all clustered at the low end of the kick range.  This is correlated with the greater degree of sphericity of models that explode early, before the post-shock turbulence can grow to significant strength and Mach number to manifest the growth of large bubbles, that when explosion ensues lead it and set an explosion direction.  For these initially non-rotating progenitors, the direction of explosion is random, but once determined a crude axis is set.  This is not the case for the lowest compactness progenitors that explode early and, hence, more spherically.

\begin{figure}
    \centering
    \includegraphics[width=0.48\textwidth]{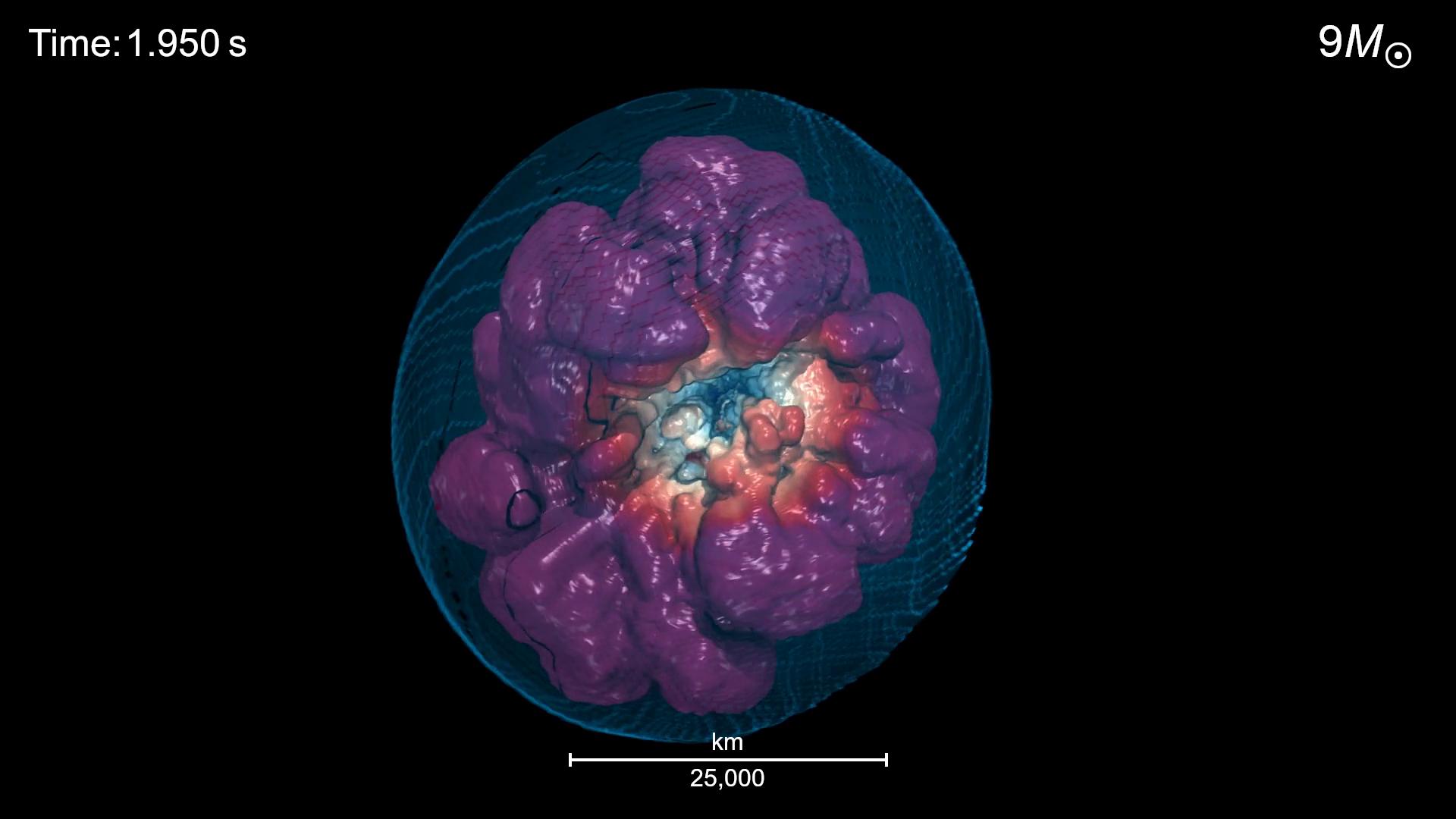}
    \includegraphics[width=0.48\textwidth]{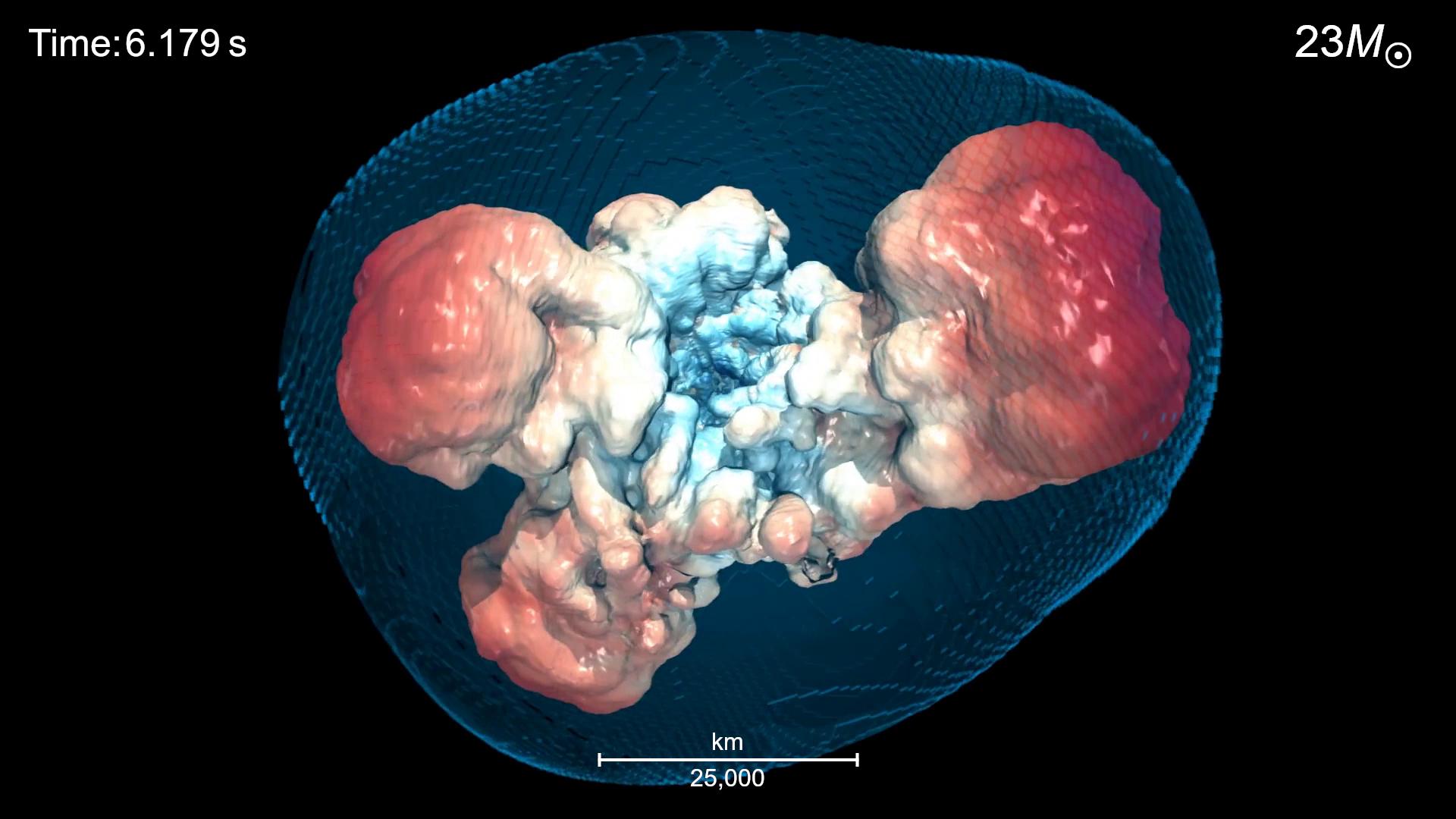}
    \caption{An isosurface (10\%) of the $^{56}$Ni abundance in the explosion of a 9 $M_{\odot}$ (left) and 23 $M_{\odot}$ (right) progenitor model at $\sim$1.95 and $\sim$6.2 seconds, respectively, after bounce, painted by Mach number.  Red/purple is positive and white/blue is negative, indicating infall. This figure is a representative depiction of a generic feature of core-collapse explosions by the turbulence-aided neutrino-driven mechanism, mainly that there is often simultaneous explosion in one general direction and accretion in a very roughly perpendicular direction when the explosion time is delayed by a few hundred milliseconds (as for the 23 $M_{\odot}$ model) and a more spherical explosion when it is launched early (within $\sim$50$-$100 milliseocnds), as for an initially non-rotating (or slowly rotating), low-compactness progenitor, such as this 9 $M_{\odot}$ model.  This fact has a bearing on the morphology of the debris, line profiles in supernova ejecta, and {possible} spin-kick correlations.  Since most explosions are delayed and manifest some bubble-driven dipolarity, the explosion direction for most progenitors (anti-)correlates with the kick direction. See text for discussions on this finding.}
    \label{fig:mach}
\end{figure}

Figure \ref{fig:mach} depicts this dichotomy.  Its left-hand side portrays the blast as traced by the 10\% iso-$^{56}$Ni
surface of the explosion of the 9 $M_{\odot}$ progenitor, colored by Mach number.  Purple is positive (outward) and blue is inward (infall).  The shock wave is the spherical bluish veil around the $^{56}$Ni bubbles.  In contrast, the right-hand side of Figure \ref{fig:mach} depicts the corresponding plot for the exploding 23 $M_{\odot}$ model \footnote{We note that the dipole and the monopole decomposition of the shock shape are unstable under roughly similar conditions at roughly the same times \citep{Burrows2013,Dolence_2013}, rendering a roughly dipolar explosion a common feature if turbulence has been allowed to grow after the shock initially stalls. This is the case for all but the quickest explosions (generally associated with the lowest compactness).}.
This model has been carried to $\sim$6.2 seconds after bounce and is only slowly asymptoting, while the 9 $M_{\odot}$ progenitor flattened within 2 seconds \citep{coleman}. The blast structure of the 23 $M_{\odot}$ model is more generic, and this is reflected in the separation of the blue dots on the right-hand side of Figure \ref{fig:total-kick-t} from the rest.  The latter are spread between $\sim$300 km s$^{-1}$ and $\sim$1000 km s$^{-1}$ and the former are found near $\sim$100$-$200 km s$^{-1}$; the difference reflects the dichotomy illustrated in Figure \ref{fig:mach}
\footnote{We emphasize that recoil kicks are merely a manifestation of momentum conservation.  Matter is exploded predominantly in one general direction over a period of time, and the residual neutron star recoils in response. This is like the rocket effect.  The effect of the gravitational accelerations is transient.  The gravitational tug of the bulk of the inner ejecta is modest and is in the wrong direction to explain the kick. There is a transient gravitational tug of matter accreting roughly opposite to the mean direction of the ejecta and this is in the general direction into which the neutron star is eventually kicked.  This matter is mostly infall/fallback and most of this matter is accreted at later times onto the final neutron star.  Hence, since these two structures eventually merge, their mutual tug effect is completely cancelled for the final, slightly fattened neutron star. As a result, the suggestion that a “gravitational tugboat” is responsible for the final kick is mostly an artifact of the failure to carry a calculation until accelerations cease and the kick asymptotes}.  

{Binary neutron stars are typically found to be on average on the low-mass side \citep{fan2023} and they seem a distinct population. One explanation is that the lower-mass neutron stars are birthed (on average) with lower kick speeds so that such binary systems can survive. This correlation emerges naturally, without tweaking, from our simulations and is consistent with observations \citep{tauris2017}. Such a  correlation was anticipated by \citet{muller_low_kick2019} and explored empirically by \citet{bray2016} and \citet{bray2018}. The right-hand side of Figure \ref{fig:total-kick-t} confirms this correlation, and also suggests that there is a neutron star mass threshold, roughly around 1.35 $M_\odot$, above which the kick velocities show much larger variations.}
It is also the case that even lower-mass progenitors not in our sample (such as the 8.8 $M_{\odot}$ progenitor of \citet{Nomoto1984} and accretion-induced collapse (AIC) systems) would naturally experience small kicks and these systems could play a role. Such progenitors are also likely to explode quickly and experience only very modest kicks (though if rapidly rotating could experience asymmetric MHD jets that could result in interesting recoils). The lowest progenitor mass range and AIC systems might be relevant as sources for the neutron stars bound in globular clusters, for which the kick speeds need to be smaller that the potential well depths ($\le$50 km s$^{-1}$). 
{Though we believe the qualitative picture is now clear, we are still some distance from a fully quantitative explanation of the binary neutron star conundrum, as suggested in population synthesis studies. More 3D numerical simulations are required to generate a complete quantitative theory that can fruitfully be compared to the observed mass-kick distribution of neutron stars. Moreover, we emphasize that the progenitor models we use in this work are all single star models \citep{Sukhbold2016,Sukhbold2018}, and that binary effects may significantly influence progenitor structures and, hence, kick velocities \citep{podsiadlowski2004}.}

\begin{figure}
    \centering
    \includegraphics[width=0.48\textwidth]{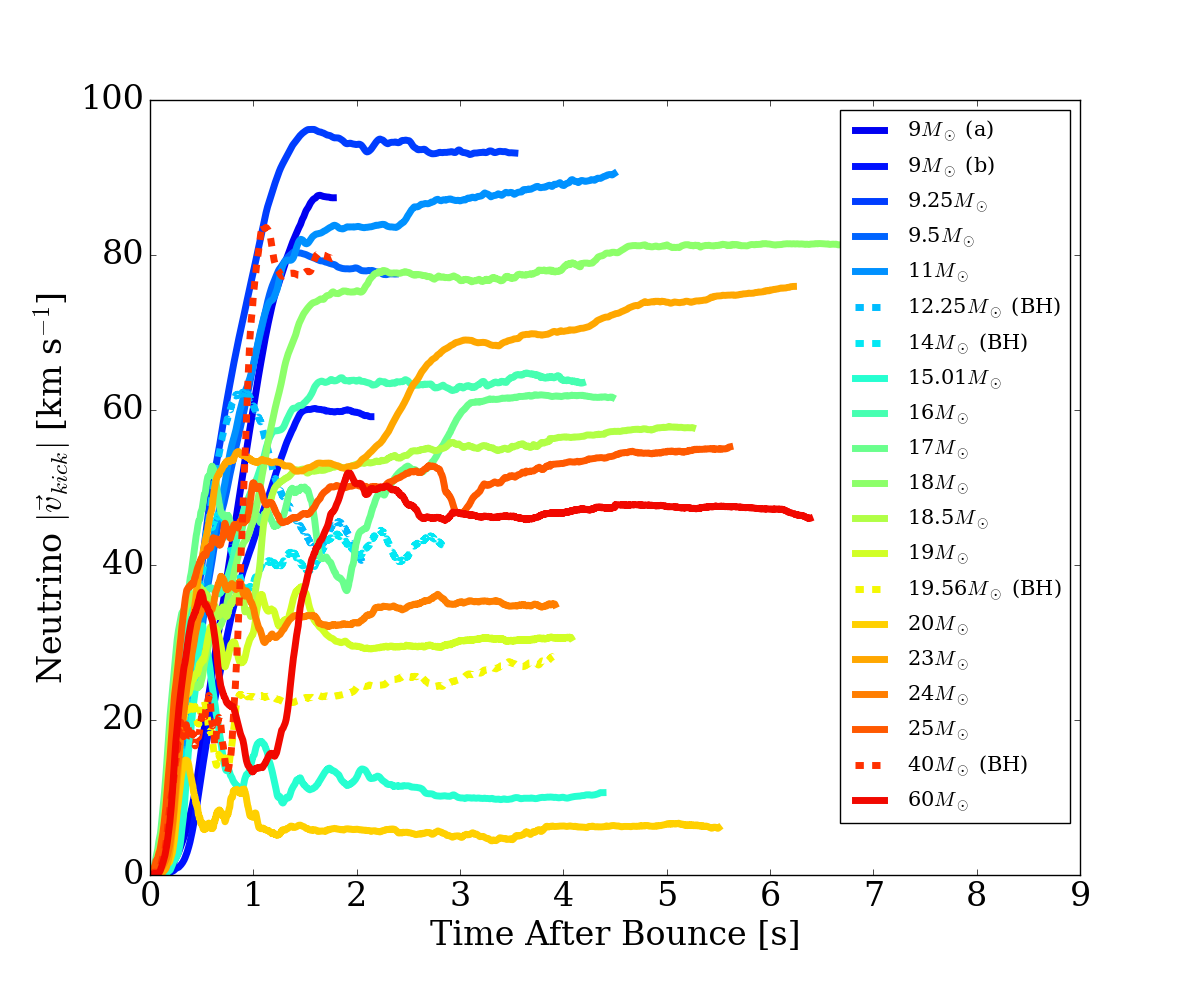}
    \includegraphics[width=0.48\textwidth]{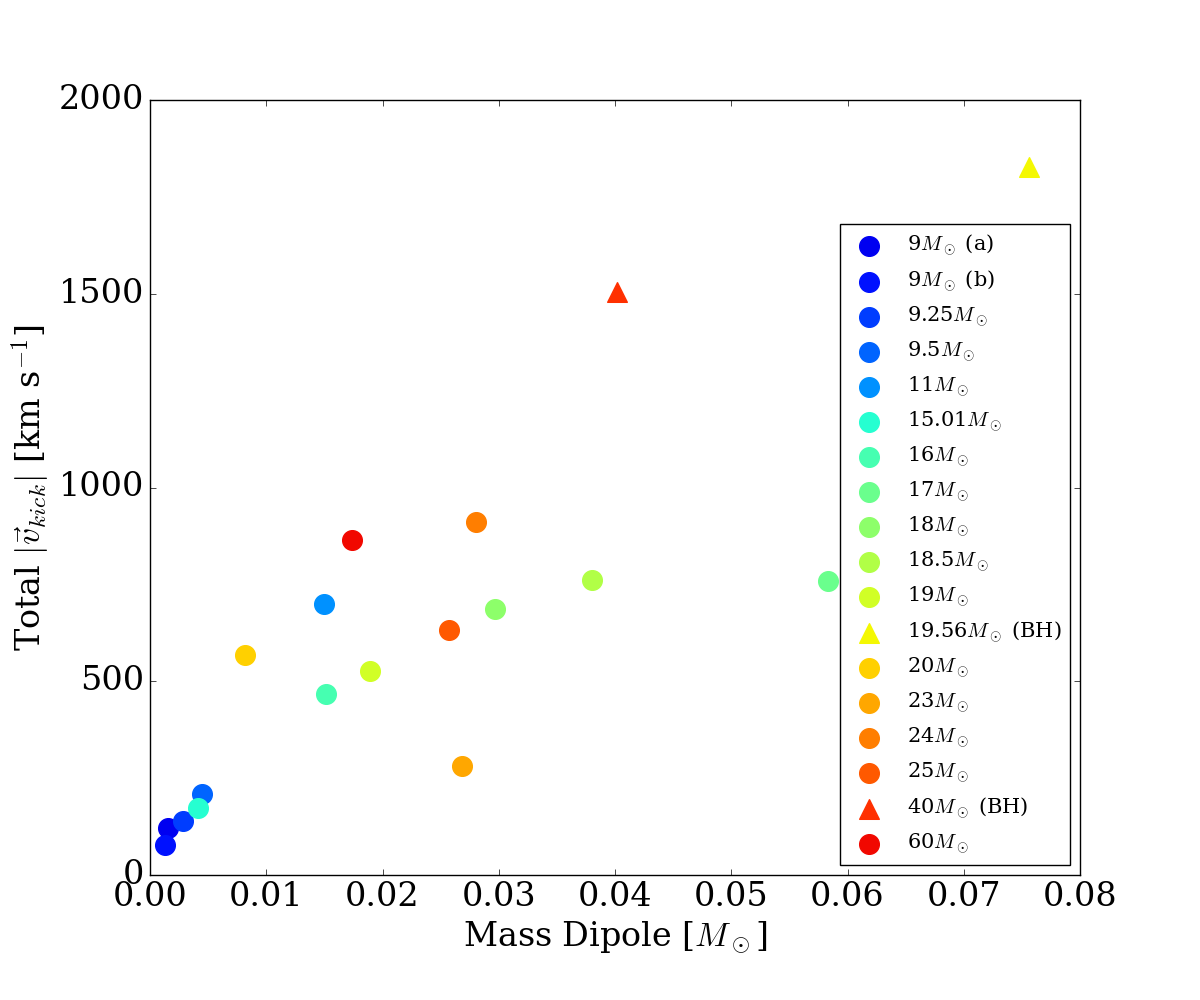}
    \caption{{\bf Left:} Neutrino kick velocity as a function of time. Solid lines are models that form neutron stars, while dashed lines are black hole formers. {\bf Right:} Mass dipole versus final total kick speed. Note that those models that seem to deviate from linearity (e.g., the 23 $M_{\odot}$ model) are also those models which have yet to asymptote, suggesting that a roughly linear relationship is even more robust than this figure implies.}
    \label{fig:nukick-t}
\end{figure}

The left-hand side of Figure \ref{fig:nukick-t} depicts 
the temporal evolution of the corresponding neutrino contribution to the kick.  This is its absolute value and in the total kick provided in Figure \ref{fig:total-kick-t}
the matter and neutrino terms are vectorially added; the two are not generally aligned and their individual instantaneous directions vary with time.  We see that the lowest ZAMS mass
progenitors (such as the 9 $M_{\odot}$ model) have some of the highest neutrino kicks \citep{coleman}, but that most other models at higher compactness whose explosions are more delayed have lower neutrino kick contributions.  Moreover, since the matter contributions to the kicks for the more delayed explosions are generally larger, the neutrino contributions are proportionately less for them and can be as low as $\sim$3\%.  They are, on the other hand, as high as $\sim$80\% for the lowest mass/compactness progenitors. We emphasize, however, that the chaos of the CCSN phenomenon can result in a range of values for all observables and that this range has not yet been well determined.

\begin{figure}
    \centering
    \includegraphics[width=0.70\textwidth]{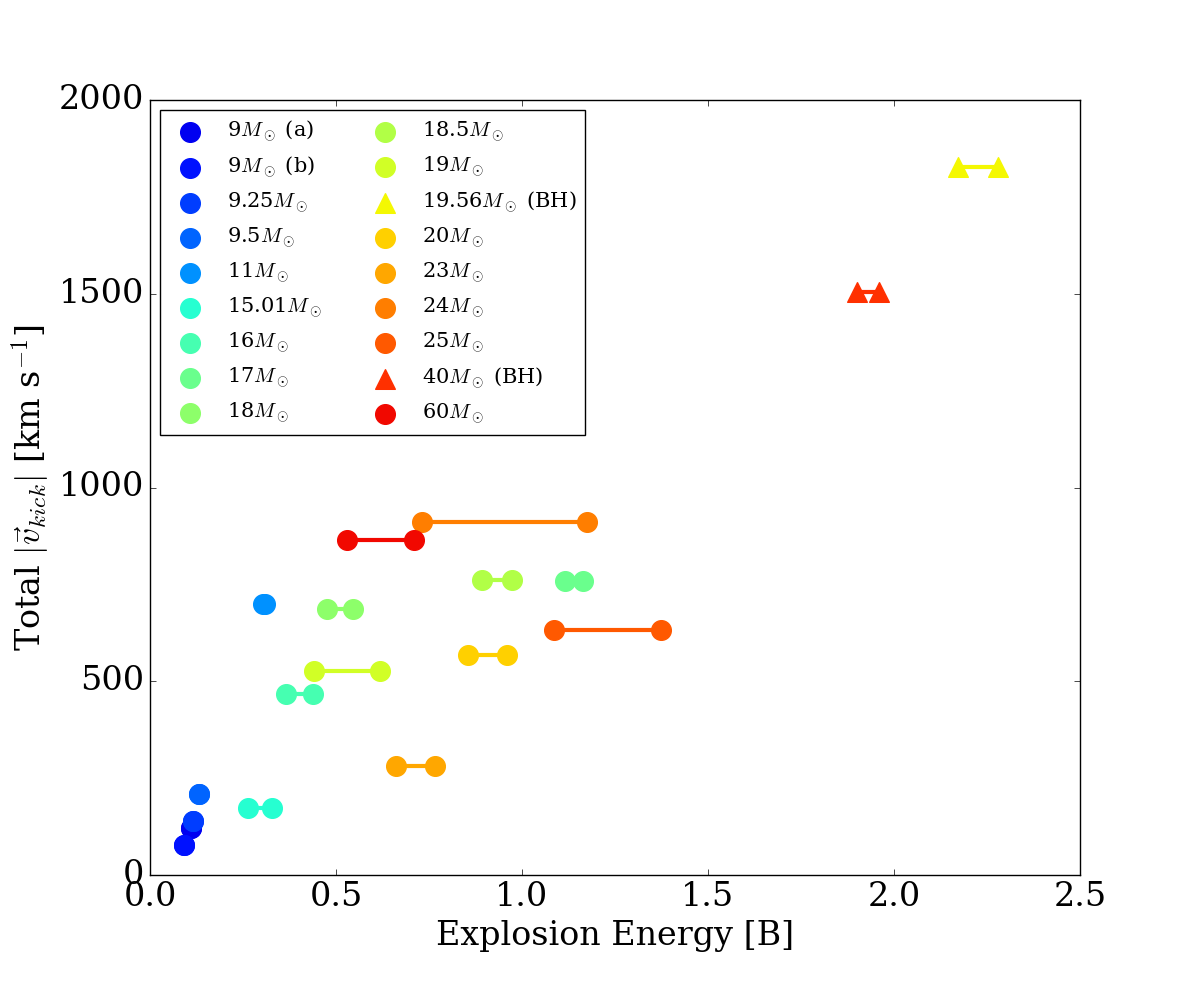}
    \caption{Magnitude of the final kick speed versus explosion energy. The range of energies reflects the fact that some of these model energies have not yet asymptoted; we provide an informed  range when they don't and the simulation of these models is still proceeding.}
    \label{fig:E-kick}
\end{figure}

On the right-hand side of Figure \ref{fig:nukick-t} we provide a plot of total kick magnitude derived versus the mass dipole of the ejecta.  The latter is a measure of the explosion asymmetry.  We see that the two are roughly correlated, as expected for the general recoil model of kicks, but that there is a bit of scatter. In Figure \ref{fig:E-kick}, we provide a similar plot of the kick magnitude versus the explosion energy.  The rough trends of kick with explosion energy and with ejecta mass dipole is clear and generally expected \citep{Burrows2007,Janka2017}.

%include 8.8, 9.6 (?), 8.4 (?) Radice; Mg = 1.19 Msun

\subsection{Black Hole Kicks}
\label{black}

Which progenitor massive stars leave stellar-mass black holes at their endpoints is not yet firm.  Despite years of speculation, 
most of which was uninformed by what at any time was the current (though evolving) state-of-the-art supernova theory, there remains no consensus among workers in the field concerning the progenitor/black-hole mapping.  The simple idea that above a ZAMS mass cut black holes are the residues of collapse and below that cut neutron stars are left now seems untenable. We ourselves have found (though not proven) that a 60 $M_{\odot}$ star leaves a neutron star (see Table \ref{tab:sims}). This is consonant with the discovery in the young cluster Westerlund 1 of a neutron star \citep{muno} that from the cluster's turnoff age must be from a progenitor with a ZAMS mass greater than 40 $M_{\odot}$. This also calls into question a helium core mass cut (perhaps near $\sim$10 $M_{\odot}$) \citep{Burrows_PT}, above which black holes form, despite the fact that the stellar-mass black-hole mass function seems to peak near this mass \citep{macleod} and such solar-metallicity stars should often leave stripped-envelope structures (either by mass transfer or winds).  

Moreover, a simplistic compactness cut, above which black holes are birthed at stellar death, is also inconsistent with the emerging theory of a turbulence-aided neutrino-driven supernova explosion. Neutrino-driven explosions with energies near $\sim$10$^{51}$ ergs require high compactness to achieve. Without rapid initial rotation and an associated MHD-driven explosion, the low compactness progenitors are not capable of generating explosions in the 10$^{51}$ ergs (one Bethe) range.  Higher compactness is necessary to provide the higher accretion luminosities and higher ``optical" neutrino depths in the absorbing post-shock mantle and gain region \citep{Bethe1985} to provide the power to drive the explosion much above the few $\times$10$^{50}$ ergs we find for the lowest mass progenitors (8.8 to 9.5 $M_{\odot}$), which possess the lowest compactnesses.

Therefore, a simple mass or compactness cut, above which black holes are the endpoints and below which neutron stars are born doesn't seem correct. This leaves this issue unresolved.  However, we \citep{wang2022,tsang2022,Burrows2020,Burrows2023} have recently found a region of progenitor space between $\sim$12 and 15 $M_{\odot}$ where for some subset of initial density profiles in that interval we do not see explosions via the neutrino mechanism using the collection of solar-metallicity massive stars above 9 $M_{\odot}$ \citep{Sukhbold2016,Sukhbold2018}. A fraction in this gap skirt explosion in the context of our 2D and 3D simulations, and this fraction could be $\sim$one-third to $\sim$two-thirds. This outcome could be because 1) the progenitor models we have employed have some flaw or because 2) our simulations may have some flaw or do not capture some essential physics. 
Nevertheless, we emphasize that in the context of our detailed 3D simulations the mass interval for this channel is almost wholly a function of the progenitor model suite, which is still very much provisional. Which mass range gives birth to black holes quiescently via the route we identify might shift significantly. 

However, perhaps such an interval is indeed a nursery for stellar-mass black hole formation.  Two of the models in the second rank of Table \ref{tab:sims}, the 12.25 and 14 $M_{\odot}$ models, 
are in this interval and do not explode.  Even after $\sim$two to three seconds after bounce, their stalled shocks continue to sink in radius. Curiously, when the shock is this compact, a spiral SASI mode \citep{Blondin2007b} is excited that modulates the neutrino and gravitational-wave signatures \citep{Vartanyan2023} in a diagnostic way. Generally, it is only when the stalled shock sinks below $\sim$120 km (usually below $\sim$100 km) that any form of SASI mode emerges from under neutrino-driven convection and this usually presages a failed explosion (though see below).

As Table \ref{tab:sims} indicates, the kicks to these black holes are quite small.  Since the models don't explode, there would be no matter recoil and the kick would be due solely to anisotropic neutrino emission.  If all the mass that resides in these models were to be accreted, given the momentum imparted to the cores and the eventual accretion of the total mass of $\sim$11$-$12 $M_{\odot}$ remaining after the integrated mass loss up to collapse is factored in \citep{Sukhbold2016,Sukhbold2018}, kicks of $\sim$7$-$8 km s$^{-1}$ are expected. However, we think it quite likely that further mass loss due to possible Roche-lobe overflow, common envelope evolution, and/or impulsive mass loss near the time of collapse \citep{morozova2018,antoni_2023} may shave more mass from these progenitors before collapse, leaving $\sim$8$-$12 $M_{\odot}$ left for black holes created through this channel.  Interestingly, this mass range roughly coincides with the peak of the measured range of masses of the relevant X-ray sources \citep{macleod}.  

Note that since these models shouldn't experience the general-relativistic instability until minutes to hours later \citep{Vartanyan2023}, and since we have not simulated to those times, our models manifest some stochastic recoil (see Table \ref{tab:sims}) due to turbulence around the still accreting PNS. However, we suggest that since the near-terminal stages of massive star evolution and Type IIp light curves \citep{morozova2018} show signs of late-time mass ejection that may be aspherical, the recoil momentum from such events might find its way into the black hole.  The result could be another component to the speed with which the black hole is born. Whether this component could have a magnitude of $\sim$100$-$200 km s$^{-1}$ \citep{Mandel2016,Fragos2009,Repetto2017,atri,Mandel2021,dashwood_brown} is to be determined. Otherwise, we find that the kicks imparted to such black holes not associated with explosions are below $\sim$10 km s$^{-1}$, with some likely scatter. 

However, as we have recently published \citep{Burrows2023} (see also \citet{chan_bh_40_2018}), there may be another channel of black hole formation that is accompanied by vigorous explosions.  Despite (or because of) the fact that these models have the highest compactness in the \citet{Sukhbold2016} solar-metallicity collection, the 19.56 and 40 $M_{\odot}$ progenitors exploded vigorously with energies of $>2\times 10^{51}$ ergs and 1.9$\times$10$^{51}$ ergs, respectively, as of $\sim$4 and 21 seconds after bounce.        

Curiously, the spiral SASI that emerged for the lower-mass black hole formers that did not encourage explosion did so for their
much-higher compactness counterparts.  This is because the high compactness models also experience high accretion rates that translate into high neutrino luminosities.  This combination
of high luminosity at a time that the spiral SASI slightly pushed out the shock and enlarged the gain region was adequate to ignite explosion.  As Table \ref{tab:sims} and Figure \ref{fig:nukick-t}
show, this channel of black hole formation not only is explosive with large energies, but has very large kicks. This is due to the
anisotropic neutrino-driven jets and anisotropic accretion
during explosion, each of which contribute to significant recoil kicks \citep{Burrows2023}.  Moreover, for these models, even though the neutrino contribution is still subdominant, it reaches values of hundreds of km s$^{-1}$. By the end of these simulations, the recoil momentum achieved is the highest among our set of twenty progenitors.  With an estimated final gravitational mass of $\sim$3.5 $M_{\odot}$ (the rest is exploded away), the 40 $M_{\odot}$ progenitor is racing at $\sim$1000 km s$^{-1}$.  Such a black hole is unlikely to remain bound to a companion. 

Therefore, we suggest there are at least two channels of black hole formation $-$ one leaving black holes quiescently in the $\sim$8$-$11 $M_{\odot}$ range with low kick speeds and another leaving black holes explosively in the $\sim$2.5 $-$ 3.5 $M_{\odot}$ mass range with significant proper motions. We note that we have not studied sub-solar-metallicity models, for which the core compactness structures and wind mass loss are quite different, that we have not simulated models above 60 $M_{\odot}$, and that we have not addressed {pulsational} pair-instability models \citep{pair} likely to provide another, though rarer, black hole formation channel. {\citet{rahman2022} do address
that channel.  However, they employ a 2D flux-limited diffusion code.
Though there are several interesting results in that paper, in our experience (and acknowledged by those authors) 2D codes are compromised in determining kick magnitudes and correlations and 3D models are generally preferred.}

\section{Induced Spins}
\label{sec:inducedspins}

Though the rotation profiles of the cores of massive stars at stellar collapse are not known, we know that massive star surfaces rotate. The loss of angular momentum to winds is guaranteed, but its magnitude is a function of surface B-fields, which themselves are less well-known. Furthermore, the redistribution of angular momentum in the stellar interior is no doubt via magnetic torques, but the processes and couplings are not firmly constrained.  The fact that red giant interiors are rotating more slowly \citep{Cantiello} than inferred using a Taylor-Spruit dynamo \citep{spruit2002} suggests a more efficient spin-down process than employed by \citet{Heger2005} in estimating the spin at collapse of massive-star models. The latter found using a Taylor-Spruit prescription for angular momentum redistribution that the Chandrasekhar cores might have initial periods at collapse near $\sim$30$-$60 seconds, which would translate by angular momentum conservation into neutron star periods of $\sim$30$-$60 milliseconds \citep{Ott2006}, with the lowest mass massive stars spinning the slowest. Hence, since these periods are not much different from those expected for radio pulsars \citep{Chevalier1986,Lyne1994,Manchester2005,Faucher2006,Popov2012,Igoshev2013,Noutsos2013}, it would seem that the issue of the initial spin period of the cores of massive stars is resolved, at least in broad outline.

However, it is still possible that many of the cores of massive stars destined to undergo collapse may be torqued down  significantly. This thought is motivated not only by the results of \citet{Cantiello}, but by observation that white dwarfs are born rotating very slowly \citep{kawaler2015,fuller_2019}.  Could this be the fate of a good fraction of the white dwarf cores of massive stars when they die? In order to explore this possibility, one first needs theoretical insight into what induced spins are possible in the context of core-collapse supernovae. This is what we provide in this paper.

\begin{figure}
    \centering
    \includegraphics[width=0.48\textwidth]{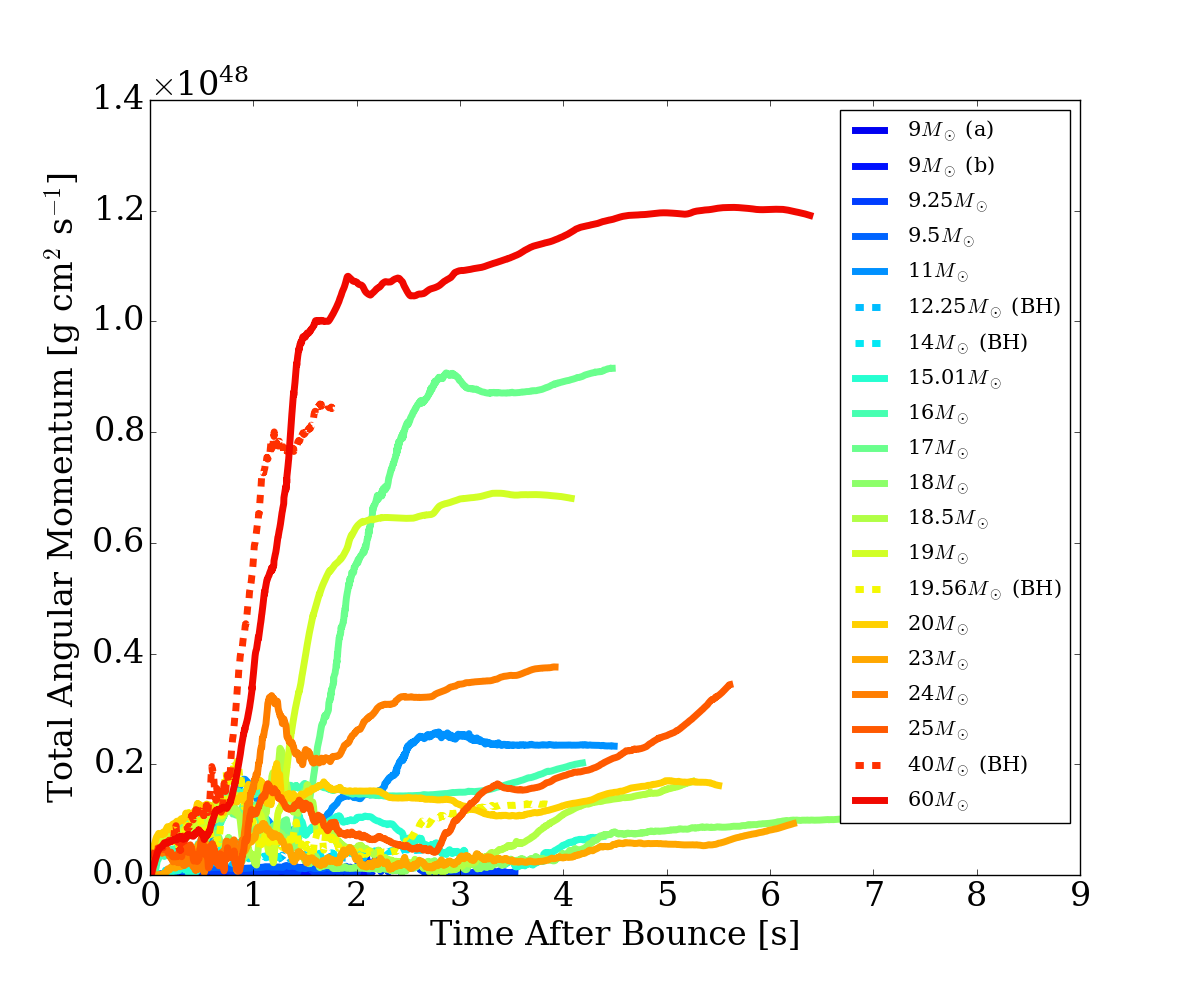}
    \includegraphics[width=0.48\textwidth]{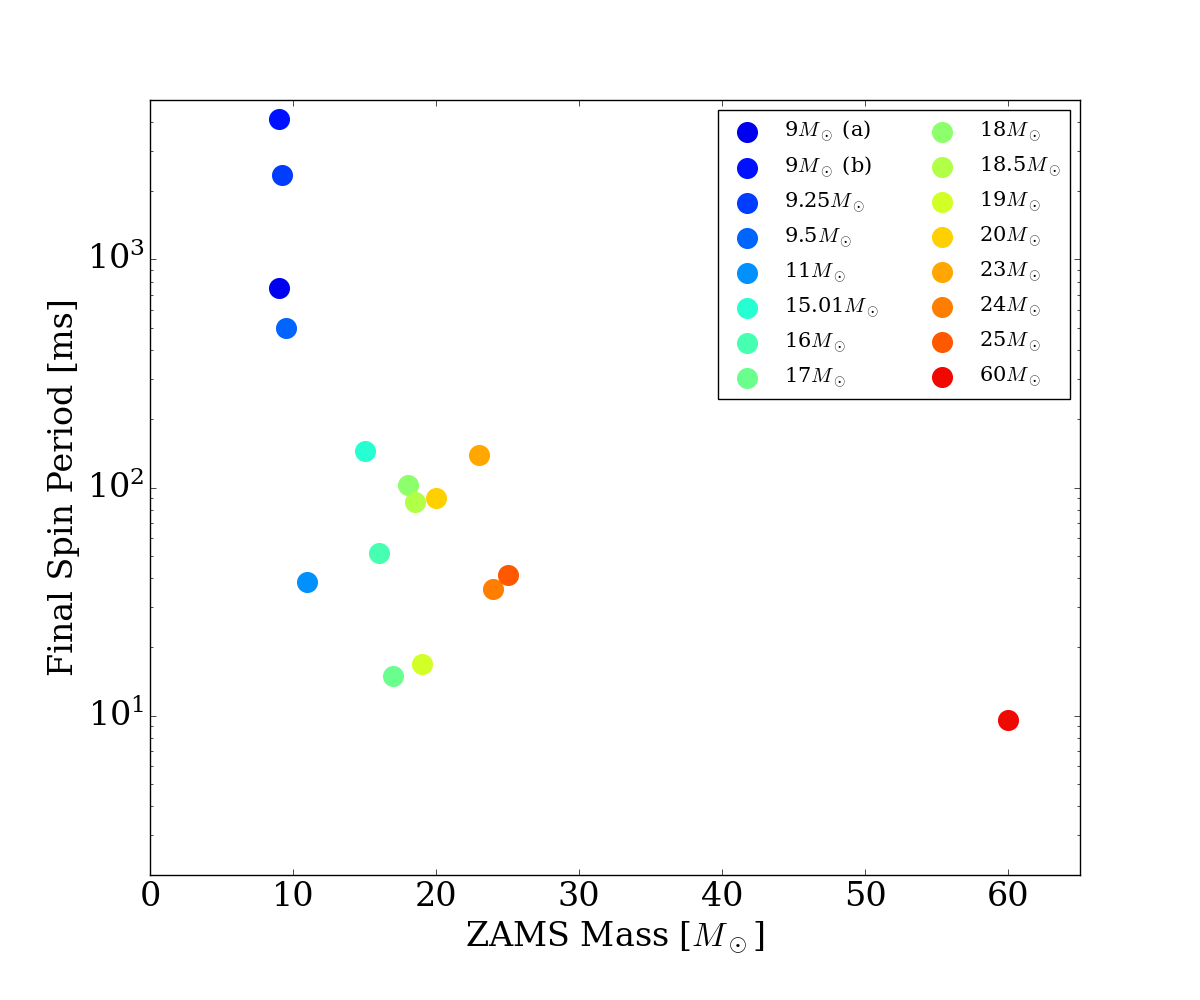}
    \caption{{\bf Left:} Total angular momentum as a function of time for each simulation. Solid lines are models that form neutron stars, while dashed lines are black hole formers. {\bf Right:} Estimated neutron star final spin period versus progenitor ZAMS mass calculated assuming $\frac{L_{tot}}{I_{NS}}$. Only neutron star formers are shown here. The neutron star moment of inertia is estimated using $(0.237+0.674\xi+4.48\xi^4)MR_{\text{NS}}^2$, where $\xi=\frac{GM}{R_{\text{NS}}c^2}$ is the compactness parameter of the neutron star itself \citep{breu2016}. We assume here that the final neutron star radius is $R_{\text{NS}}=12$ km.}
    \label{fig:L-t}
\end{figure}

The left-hand side of Figure \ref{fig:L-t} renders the evolution
of the angular momentum imparted to the PNS cores during their post-bounce evolution. This spin up is due to the stochastic accretion of plumes of infalling matter into the PNS, with varying impact parameters.  If the explosion is very asymmetrical, as is the case when the kick speeds are significant, or black hole formation is accompanied by an explosion, the consequently asymmetrical accretion during explosion (mostly then from one side) can spin the PNS up to interesting periods. For the black holes that form in an explosion, these spins can reach high values (see Table \ref{tab:sims}), even though their cores were initially not rotating.  This implies that such objects could be collapsar candidates \citep{Heger2003}, despite their lack of initial spin and their solar-metallicity progenitors. 

The right-hand side of Figure \ref{fig:L-t} portrays the dependence of the final induced spin period of the neutron star upon ZAMS mass. We see a trend, roughly recapitulated with the kick systematics (see Table \ref{tab:sims}), of decreasing period with increasing ZAMS mass, but with a lot of scatter.  Such scatter should be expected in the context of the chaotic turbulence seen in 3D simulations of CCSNe.  We emphasize that the lowest ZAMS mass and compactness progenitors which explode early and approximately spherically have the slowest induced spin rates, while those exploding later and at higher compactness have greater induced spin rates.  In fact, our highest compactness models can leave cold neutron stars with high angular momenta and spins. However, while we are comfortable with the compactness/kick and PNS mass/kick correlations we derive, we are much less sanguine that induced spins are the whole story of compact object birth spins. Importantly, we know that the Crab explosion involved a low-mass progenitor near $\sim$9$-$10 $M_{\odot}$ \citep{Stockinger2020}, but that its birth spin was $\sim$15 milliseconds \citep{Lyne2015}. This is far from the induced periods (many seconds) we find for this mass range of progenitors, not only suggesting, but requiring, the story to be more complicated.  Spin-up by binary interaction is also distinctly possible \citep{Fuller2022}. Therefore, the relative roles of induced and initial spins in the distribution of observed birth periods for compact objects remains open. 

\begin{figure}
    \centering
    \includegraphics[width=0.48\textwidth]{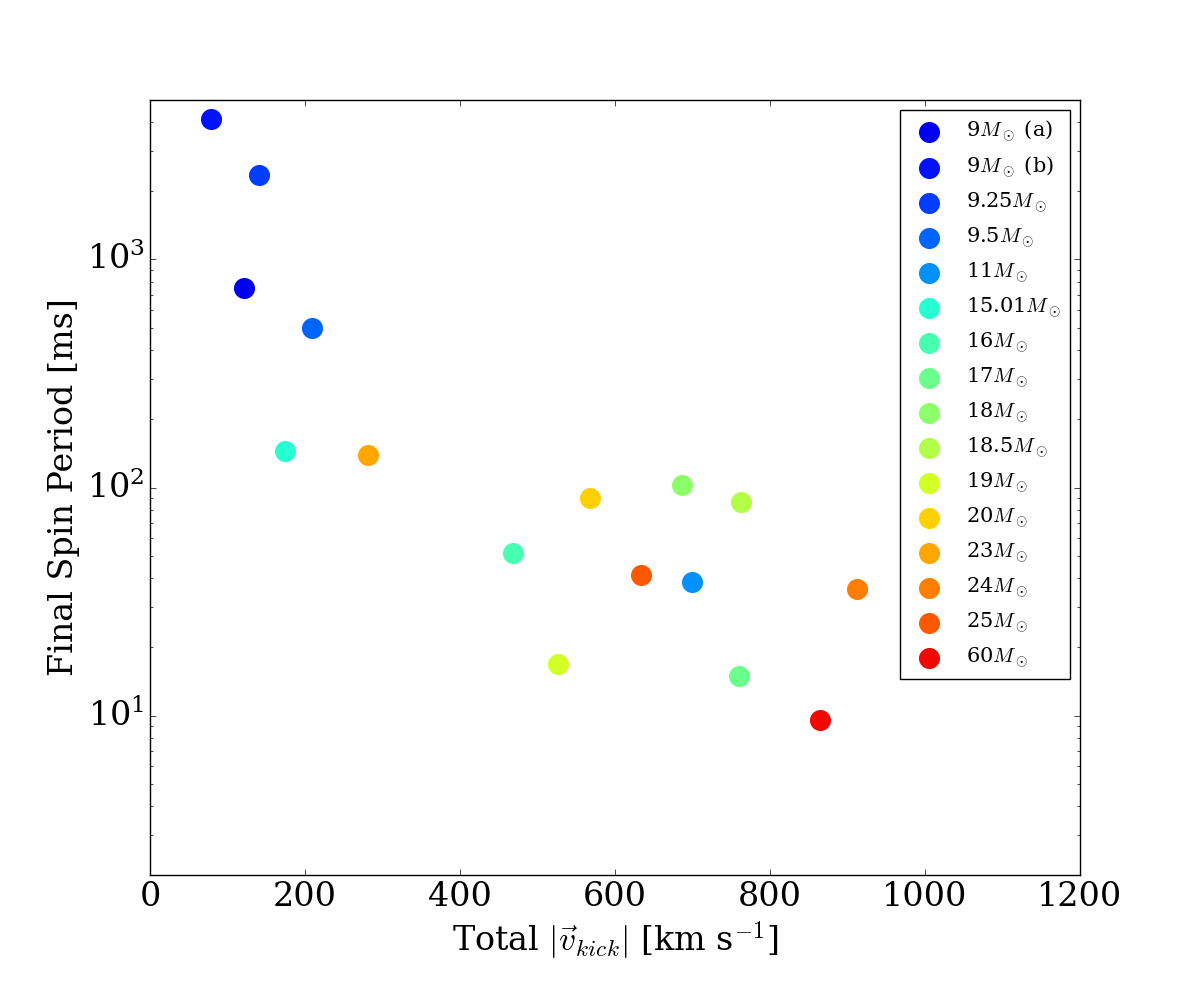}
    \caption{Relation between kick speeds and induced spins of neutron stars birthed in initially non-rotating progenitor stars. See text for a discussion.}
    \label{fig:kick-vs-spin}
\end{figure}

Nevertheless, we find a weak, but intriguing, correlation between induced spin and recoil kicks, demonstrated in Figure \ref{fig:kick-vs-spin}.  The greater the kick, the greater the induced spin, again with significant scatter. Is there something of importance in this apparent correlation? Is there rough support for this, however complicated by the overlapping effect of the initial spin, in the pulsar database? 

The induced spins for the non-exploding black hole channel (see Table \ref{tab:sims}) are quite low, again resulting solely from from the torques due to the radiated neutrinos.  Spin parameters of close to zero are inferred from our simulations and model extrapolations.  However, should there be any significant angular momentum in the accreting stellar envelope that on long timescales will bulk up the core (something quite likely), these numbers can be radically changed.  Hence, even under the assumption that the initial core spin rate is zero, we can not say with any confidence what the final spin parameter of black holes born through this channel, might be.

However, the induced spins for the exploding black hole channel can be large.  This can follow from the significant asymmetry in their explosions, itself related to the large recoil kick. A spin parameter as large as $\sim$0.6 emerges for the 40 $M_{\odot}$ model (see Table \ref{tab:sims}), though we have yet to determine the corresponding number for the 19.56 $M_{\odot}$ model. Their large compactnesses and related large post-explosion asymmetrical mass accretion rates are direct causative factors. In addition, accretion continues to power the driving neutrino emissions, which for this black hole formation channel are quite large. Accretion settles into biased accretion onto one general side and spins up the periphery of the PNS before black hole formation.  After black hole formation, continued accretion parallels continued spin up until after many seconds the black hole is isolated (for the 40 $M_{\odot}$ model after $\sim$50 seconds?). This general black hole formation scenario is discussed in \citet{Burrows2023}.

\section{Spin-Kick Correlation}
\label{correlation}

\begin{figure}
    \centering
    \includegraphics[width=0.48\textwidth]{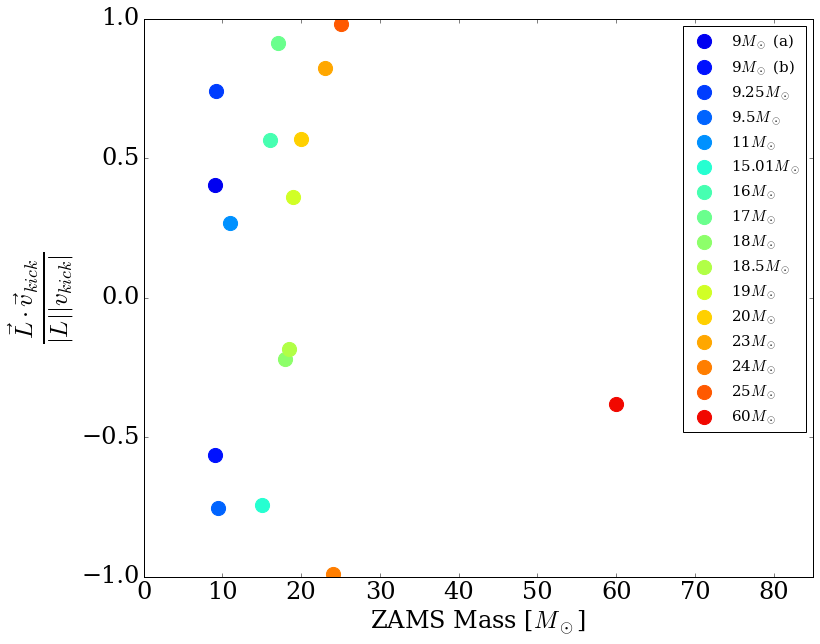}
    \includegraphics[width=0.48\textwidth]{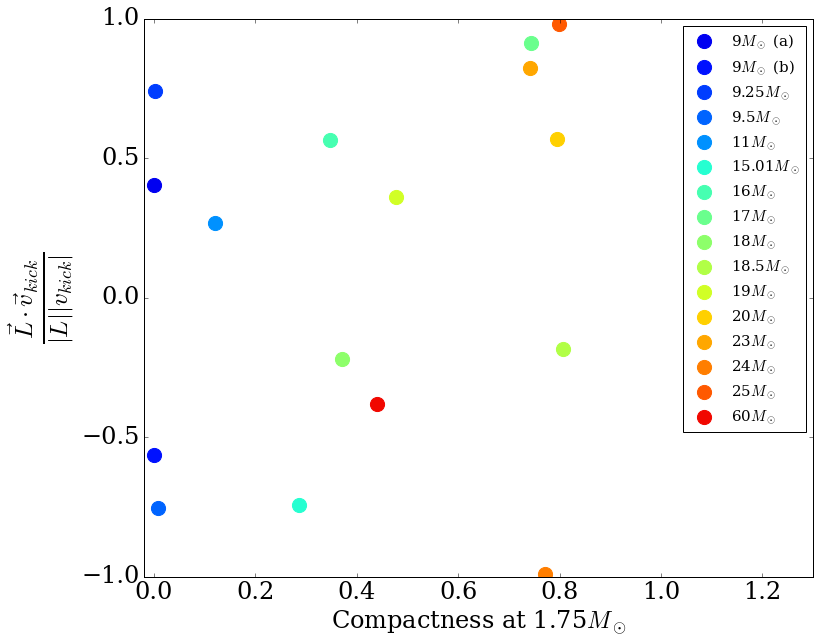}
    \caption{Cosine of the angle between the neutron star kick and induced spin directions versus ZAMS mass (left) and compactness (right).}
    \label{fig:spin-kick-angle}
\end{figure}

Figure \ref{fig:mach} portrays the Mach number for $^{56}$Ni surfaces and the crudely axial explosions when the compactness is not small and when there is a greater delay to explosion. It also indicates that there is continued accretion roughly in the perpendicular plane. Hence, when the compactnesses and ZAMS masses are not small, explosion and accretion are roughly perpendicular.  Accretion then spins up the periphery of the PNS and the net spin vector is either roughly parallel or anti-parallel to the kick (which is along or anti-parallel to the explosion axis). Hence, {we would think} a spin-kick correlation \citep{Holland-Ashford2017,2018ApJ...856...18K,Ng2007} would naturally emerge. Figure \ref{fig:spin-kick-angle} depicts the dot product of the final spin and kick unit vectors for our model set, versus ZAMS mass (left) and compactness (right). We see that {though} there seems to be a deficit in the perpendicular direction and there is a slight preference for an (anti-)aligned orientation {the null hypothesis can not be rejected}. {Indeed}, more rigorously, we perform a Kolmogorov–Smirnov (K-S) test with the null hypothesis that the simulated angles are drawn from a uniform distribution on the sphere. We see a D-statistic of 0.19, which means the cumulative distributions are not very different. In addition, due to the small number of models, the p-value of this K-S test is 0.51, which means that there is a chance that the deficit in the perpendicular direction is simply due to statistical fluctuations. Therefore, however intriguing, many more simulations are required to {properly address a potential spin-kick correlation in the initially non-rotating case.}  

Nevertheless, such a correlation may arise when the Chandrasekhar core is initially rotating, since such rotation may slightly bias (depending upon its magnitude) the axis and ejecta mass dipole vector of the subsequent neutrino-driven explosion. If such is the case, then the spin-kick correlation may {emerge}.  Continued accretion is very roughly perpendicular to the explosion direction, itself roughly (anti-) parallel to the kick direction, resulting in a (anti-)correlation between spin and kick. Note that we have not invoked B-fields nor MHD jets and that this is an expected outcome in the neutrino-driven explosion context.  However, we emphasize that we have yet to perform the requisite rotating simulations.

\section{Conclusions}
\label{conclusions}

With a focus on the recoil kicks imparted to the residues of stellar core collapse in the supernova and black hole formation contexts, we have analyzed twenty state-of-the-art 3D long-term core-collapse simulations generated using the code F{\sc{ornax}}. In the past, most 3D models were not of sufficient duration to witness the cessation of net acceleration and the asymptoting of the kicks to their final values. This includes our own previous work \citep{coleman}, as well as that of other workers in the field.  However, and for the first time for a large and uniform collection of 3D supernova models, we have either asymptoted the kicks, or come within 20\% of doing so. As a result, we obtain an integrated and wide-angle perspective of the overall dependence of the recoil kicks and induced spins upon progenitor mass and their Chandrasekhar-like core structures, the latter indexed approximately by compactness.
As had been postulated in the original recoil kick model \citep{janka1994,Burrows1996}, the recoil is always opposite in direction to the bulk of the ejecta. 

We find that lower mass and lower compactness progenitors that explode quasi-spherically due to the short delay to explosion experience on average smaller kicks in the $\sim$100$-$200 km s$^{-1}$ range, while higher mass and higher compactness progenitors that explode later and more aspherically give birth to neutron stars with kicks in the $\sim$300$-$1000 km s$^{-1}$ range.  We also find that these two classes can be correlated with the gravitational mass of the residual neutron star. The correlation of a lower kick speed with lower neutron star gravitational mass suggests that the survival of binary neutron star systems, for which their mean mass is lower than average, may in part be due to the lower kick speeds we see for them. With some scatter, there is a correlation of the kick with both the mass dipole of the ejecta and the explosion energy.  We also find, as did \citet{coleman}, that the contribution of the recoil due to anisotropic neutrino emission can be important for the lowest mass/compactness progenitors.  

However, unlike \citet{coleman}, we partition the fractional contribution of the neutrino component to the overall kick differently, which results in an apparent diminution of its contribution for higher ZAMS mass and compactness progenitors, for which the matter recoil effect then predominates. The difference with \citet{coleman} is a consequence of our use here of the standard literature radius at which to calculate the neutrino contribution, and not due to some major differences with the overall results in \citet{coleman}. Between the effective radius at which the neutrino component is calculated in \citet{coleman} and here, there is momentum transfer between the matter and neutrino radiation that is properly taken into account in the vector sum of the two provided in both papers.  

We find two channels for black hole birth.  The first leaves masses of $\sim$10 $M_{\odot}$, is not accompanied by a neutrino-driven explosion, and experiences small kicks, perhaps below $\sim$10 km s$^{-1}$. The second \citep{Burrows2023} is through the vigorous, neutrino-driven explosion of a high-compactness progenitor that leaves behind a black hole with a mass of perhaps $\sim$2.5$-$3.5 $M_{\odot}$ that is kicked to high speeds, perhaps above $\sim$1000 km s$^{-1}$. This second black-hole formation channel is novel and unexpected, but if true challenges most notions of what can explode. Both channels challenge prior notions of the mapping between progenitor and outcome and must still be verified. However, in the context of our CCSN simulations, the associated mass ranges for black hole formation depend entirely upon the as yet unconverged progenitor suite model.  In particular, the ZAMS mass range for our quiescent black hole formation channel could shift significantly when binarity, overshoot, shell-merger, mixing processes, and the $^{12}$C($\alpha$,$\gamma$) rate are properly understood.  

Associated with kicks are induced spins that accompany them in the context of the modern theory of turbulence-aided, neutrino-driven supernova explosions.  We find that the induced spins of nascent neutron stars range from a few seconds at low compactness and low progenitor mass to $\sim$10 milliseconds at higher compactness and higher ZAMS mass. However, this can not be the whole story of the birth spins of neutron stars, 
and the relative roles of initial and induced spins have yet to be determined.  Nevertheless, we find that it is in principle possible to explain the some spin rates of some pulsars with the post bounce asymmetrical accretion of net angular momentum in the context of asymmetrical supernova explosions for which spherical symmetry is generically broken.   Moreover, we find that a spin/kick correlation for pulsars {is not yet suggested by the theory}.  However, if there is an initial spin to the collapsing core, its vector direction could bias the direction of the explosion. If this is the case, a spin/kick correlation {may emerge}.  However, this speculation has yet to be verified, but, if true, a spin/kick correlation (in a statistical sense) {would} naturally emerge as a byproduct of the turbulence-aided neutrino mechanism of core-collapse supernova explosions.  

In the non-exploding case of black hole formation, the induced spins are quite low and due to anisotropic neutrino emission. Any slight neutrino torques would be due to a random stochastic residual effect.  Under these circumstances, an initial spin would be determinative. However, for the exploding context of black hole formation, the induced spin is likely to be large.  For our 40 $M_{\odot}$ progenitor, we find that the spin parameter could reach $\sim$0.6 (the maximum is 1.0).  This is in the collapsar range, and suggests that even an initially non-rotating, solar-metallicity, high compactness progenitor could be a seat of long-soft gamma-ray bursts, with a vigorous explosive precursor.  This is a controversial claim, and one that needs much further scrutiny.  However, this is what is suggested by our simulations.

It must be remembered that the chaos in the turbulence generic and fundamental to core collapse phenomena will result in a spread of outcomes, even for a given progenitor.  This translates into distribution functions for the explosion energies, nucleosynthesis, morphology, kicks, and induced spins that are not delta functions and are currently unknown.  Hence, there is natural scatter in the kicks and induced spins for a given initial model that has yet to be determined, but that will spread observed values.  Some of the scatter seen in this study no doubt originates in this behavior. Such are the wages of chaos. Nevertheless, though the study of a single 3D model could distort one's understanding, the study of a wide range of progenitor structures should provide a
more complete picture of the theoretical trends and correlations that will minimize some of the confusion due to such scatter.  In any case, this is the philosophy of this 3D multi-model, long-term exploration of the core-collapse, but much more remains to be done. In addition, we note that the mapping of observables to progenitor structure, modeled loosely as ``compactness," seems more robust than the mapping with ZAMS progenitor mass, tethered as it is to the progenitor suite employed. Given this, the fact that we are in this study using the Sukhbold set of progenitors limits what one can say about the observable/ZAMS-mass correlations and the guidance we can provide to population synthesis modelers. The effects of binarity (mass transfer), overshoot, shell mergers, perturbations, rotation, the $^{12}$C($\alpha$,$\gamma$) nuclear rate, and wind mass loss have yet to be fully understood and their influence retired. Moreover, physical processes that might change residual black hole properties on timescales of hours, days, or years, such as ``fallback'' \citep{schroder2018,Chan2020} or random late-time angular momentum accretion \citep{antoni_2023}, are not covered in this work.

Though F{\sc{ornax}} is a sophisticated tool, it incorporates no approach to neutrino oscillations, uses a moment-closure approach to the radiative transfer, and incorporates approximate, not precise, general relativity.  Nevertheless, the greatly expanded panorama the new capabilities and this new numerical database provide is qualitatively more comprehensive and predictive, due to its breadth and temporal extent, than has heretofore been possible and appears to explain many observed pulsar properties by default.

\section*{Acknowledgments}

We thank Chris White for previous insights, collaboration, and conversations. We acknowledge support from the U.~S.\ Department of Energy Office of Science and the Office of Advanced Scientific Computing Research via the Scientific Discovery through Advanced Computing (SciDAC4) program and Grant DE-SC0018297 (subaward 00009650) and support from the U.~S.\ National Science Foundation (NSF) under Grants AST-1714267 and PHY-1804048 (the latter via the Max-Planck/Princeton Center (MPPC) for Plasma Physics). 
Some of the models were simulated on the Frontera cluster (under awards AST20020 and AST21003), and this research is part of the Frontera computing project at the Texas Advanced Computing Center \citep{Stanzione2020}. Frontera is made possible by NSF award OAC-1818253. Additionally, a generous award of computer time was provided by the INCITE program, enabling this research to use resources of the Argonne Leadership Computing Facility, a DOE Office of Science User Facility supported under Contract DE-AC02-06CH11357. Finally, the authors acknowledge computational resources provided by the high-performance computer center at Princeton University, which is jointly supported by the Princeton Institute for Computational Science and Engineering (PICSciE) and the Princeton University Office of Information Technology, and our continuing allocation at the National Energy Research Scientific Computing Center (NERSC), which is supported by the Office of Science of the U.~S.\ Department of Energy under contract DE-AC03-76SF00098.

%%%%%%%%%%%%%%%%%%%%%%%%%%%%%%%%%%%%%%%%%%%%%%%%%%
\section*{Data Availability}

The numerical data underlying this article will be shared upon reasonable request to the corresponding author.

%%%%%%%%%%%%%%%%%%%% REFERENCES %%%%%%%%%%%%%%%%%%

%\nocite{*}   

%\bibliographystyle{plain}
\bibliography{citations}{}
\bibliographystyle{aasjournal}

\begin{sidewaystable}
%\centering
\begin{tabular}{c|ccccccccccc}
\hline
$M_\text{ZAMS}$ & $\xi_{1.75}$       & $t_\text{max}$ & Type   & $M_\text{baryonic}$ & $M_\text{grav}$  & $p_\text{kick}/10^{40}$ & $v_\text{kick}^\text{total}$ & $v_\text{kick}^\nu$ & $L/10^{46}$             & $P_\text{cold}$ &$\hat{L}\cdot\hat{v}$\\
{[}$M_\odot${]} &                    & {[}s{]}        &        & {[}$M_\odot${]}     & {[}$M_\odot${]}  & {[}g cm s$^{-1}${]}     & {[}km s$^{-1}${]}            & {[}km s$^{-1}${]}   & {[}g cm$^2$ s$^{-1}${]} & {[}ms{]}     &   \\ \hline
9(a)            & $6.7\times10^{-5}$ & 1.775          & NS     & 1.347               & 1.237            & 3.235                   & 120.7                        & 87.4                & 1.046                   & 749.1        &0.402   \\
9(b)            & $6.7\times10^{-5}$ & 2.139          & NS     & 1.348               & 1.238            & 2.106                   & 78.6                         & 59.1                & 0.190                   & 4132         &-0.564   \\
9.25            & $2.5\times10^{-3}$ & 3.532          & NS     & 1.378               & 1.263            & 3.842                   & 140.1                        & 93.2                & 0.344                   & 2346         &0.742   \\
9.5             & $8.5\times10^{-3}$ & 2.375          & NS     & 1.397               & 1.278            & 5.796                   & 208.6                        & 77.5                & 1.636                   & 501.5        &-0.756   \\
11              & 0.12               & 4.492          & NS     & 1.497               & 1.361            & 20.83                   & 699.4                        & 90.5                & 23.28                   & 38.54        &0.266   \\
15.01           & 0.29               & 4.384          & NS     & 1.638               & 1.474            & 5.668                   & 173.9                        & 10.6                & 6.995                   & 144.2        &-0.743   \\
16              & 0.35               & 4.184          & NS     & 1.678               & 1.505            & 15.62                   & 468.0                        & 63.5                & 20.23                   & 51.45        &0.566   \\
17              & 0.74               & 4.473          & NS     & 2.044               & 1.785            & 30.90                   & 759.8                        & 61.6                & 91.52                   & 14.9        &0.914   \\
18              & 0.37               & 6.778          & NS     & 1.684               & 1.510            & 22.97                   & 686.0                        & 81.4                & 10.23                   & 102.2        &-0.221   \\
18.5            & 0.80               & 5.250          & NS     & 2.139               & 1.854            & 32.43                   & 762.4                        & 57.7                & 16.88                   & 86.14        &-0.184   \\
19              & 0.48               & 4.075          & NS     & 1.803               & 1.603            & 18.89                   & 526.7                        & 30.6                & 68.05                   & 16.85        &0.360   \\
20              & 0.79               & 5.503          & NS     & 2.143               & 1.857            & 24.21                   & 567.9                        & 6.08                & 16.17                   & 90.22        &0.568   \\
23              & 0.74               & 6.228          & NS     & 1.959               & 1.722            & 10.94                   & 280.8                        & 75.9                & 9.294                   & 138.2        &0.825   \\
24              & 0.77               & 3.919          & NS     & 2.028               & 1.773            & 36.79                   & 912.1                        & 34.9                & 37.58                   & 35.88        &-0.990   \\
25              & 0.80               & 5.611         & NS     & 2.106               & 1.830             & 26.52                   & 633.0                        & 55.2                & 34.38                   & 41.37        &0.979   \\
60              & 0.44               & 6.386          & NS     & 1.791               & 1.594            & 30.81                   & 864.8                        & 46.0                & 119.1                   & 9.537        &-0.380   \\ \hline
$M_\text{ZAMS}$ & $\xi_{1.75}$       & $t_\text{max}$ & Type   & $M_\text{baryonic}$ & $M_\text{final}$ & $p_\text{kick}/10^{40}$ & $v_\text{kick}^\text{total}$ & $v_\text{kick}^\nu$ & $L/10^{46}$             & $a$          &$\hat{L}\cdot\hat{v}$   \\
{[}$M_\odot${]} &                    & {[}s{]}        &        & {[}$M_\odot${]}     & {[}$M_\odot${]}  & {[}g cm s$^{-1}${]}     & {[}km s$^{-1}${]}            & {[}km s$^{-1}${]}   & {[}g cm$^2$ s$^{-1}${]} &                 \\ \hline
 12.25           & 0.34               & 2.090          & BH (F) & 1.815               & $\sim$11.1       & 1.441                   & 39.9 (10.7)                  & 39.9 (6.90)         & 3.621                   & 0.002 (0.0004)  & -0.677\\
14              & 0.48               & 2.824          & BH (F) & 1.990               & $\sim$12.1       & 1.684                    & 42.5 (9.59)                  & 42.5 (7.9)          & 4.638                   & 0.003 (0.0004)  &-0.920\\
19.56           & 0.85               & 3.890          & BH (S) & 2.278               & ?                & 82.76                   & 1826 (?)                     & 28.1 (?)            & 12.83                   & 0.006 (?)    &0.966   \\
40              & 0.87               & 1.76 (21)          & BH (S) & 2.434               & $\sim$3.5        & 72.85                   & 1504(1034)                   & 79.5 (83.0)         & 84.32                   & 0.039 (0.6)  &-0.883      
\end{tabular}
\caption{
    Results for the simulations presented in this work. 
    $M_{\rm ZAMS}$ is the initial ZAMS mass of the progenitor model.
    $t_{\rm max}$ is the maximum post-bounce time reached by the simulation. For the black hole formers, the times are the time of black hole formation, and the second time (if shown) is the latest time after black hole formation we have carried that model hydrodynamically. The types of central objects are summarized. NS means neutron star, BH (F) means black hole formed with a failed explosion, and BH (S) means black hole formed accompanied by a successful CCSN explosion.
    $M_{\rm baryonic}$ is the baryonic mass of the central object at $t_{\rm max}$, while $M_{\rm grav}$ is the gravitational mass of a cold, catalyzed NS. $M_{\rm final}$ is the estimated final black hole mass. The final masses of the non-exploding black hole formers are the total masses in the progenitor models at collapse, assuming all matter will eventually be accreted, while the final masses of the exploding black hole formers are derived from our hydro-only blast wave simulations after black hole formation.
    $p_{\rm kick}$ gives the linear momentum of the object at $t_{\rm max}$.
    $v_{\rm kick}^{\rm(total)}$ and $v_{\rm kick}^{\rm(\nu)}$ are the total kick speeds and the kick speeds induced by neutrinos only, respectively. Both are measured at $t_{\rm max}$. Numbers in brackets show the estimated final values (kick speed, spin parameter) of the black holes. Note that the 40-$M_{\odot}$ model has been simulated to 21 seconds after bounce, at which point the spin parameter has reached $\sim$0.35, and that we estimate its final gravitational mass and spin parameter ($a$) will be $\sim$3.5 $M_{\odot}$ 0.6, respectively, achieved after a further $\sim$25 seconds of evolution. We have yet to determine the corresponding numbers for the 19.56 $M_{\odot}$ model. These numbers still need to be checked. $L_{\rm central}$ is the angular momentum of the central object at $t_{\rm max}$, and
    $P_{\rm cold}$ is the inferred final spin period of the cold final neutron star, assuming $P=\frac{2\pi L_\text{central}}{I_\text{NS}}$. The neutron star moment of inertia is estimated using $I_\text{NS}=(0.237+0.674\xi_{\rm NS}+4.48\xi_{\rm NS}^4)MR_{\text{NS}}^2$, where $\xi_{\rm NS}=\frac{GM}{R_{\text{NS}}c^2}$ is the compactness parameter of the neutron star itself \citep{breu2016}. We assume here that the final neutron star radius is $R_{\text{NS}}=12$km. For black holes, the relativistic spin parameter $a=\frac{Lc}{GM^2}$ is shown and the number in brackets is our guess for its of final value. The final column contains the cosine of the spin/kick relative angle at the end of each simulation.
    }
    \label{tab:sims}
\end{sidewaystable}

\clearpage

%%%%%%%%%%%%%%%%%%%%%%%%%%%%%%%%%%%%%%%%%%%%%%%%%%

%%%%%%%%%%%%%%%%%%% FIGURES %%%%%%%%%%%%%%%%%%%%%%

\end{document}